\documentclass{article}
\usepackage{spconf,amsmath,graphicx}
\usepackage{caption}
\usepackage{subcaption}
\usepackage{algorithm}
\usepackage{algpseudocode}
\usepackage{setspace}

\title{Channel-Aware Pretraining of Joint Encoder-Decoder Self-Supervised Model for Telephonic-Speech ASR}
\name{Vrunda N. Sukhadia, S. Umesh }
\address{Speech Lab, Electrical Engineering Department, IIT Madras, India}
\begin{document}
%
\maketitle
\begin{abstract}

This paper proposes a novel technique to obtain better downstream ASR performance from a joint encoder-decoder self-supervised model when trained with speech pooled from two different channels (narrow and wide band). The joint encoder-decoder self-supervised model extends the \textit{HuBERT} model with a Transformer decoder. \textit{HuBERT} performs clustering of features and predicts the class of every input frame. In simple pooling, which is our baseline, there is no way to identify the channel information. To incorporate channel information, we have proposed non-overlapping cluster IDs for speech from different channels. Our method gives a relative improvement of $\sim4\%$ over the joint encoder-decoder self-supervised model built with simple pooling of data, which serves as our baseline.

\end{abstract}

\section{Introduction}
Telephonic Automatic Speech Recognition (ASR) is one of the key ASR applications. Speech signals in Telephonic channels are generally narrow band (sampled at $\sim$ 8kHz) in nature. Wide band (sampled at $\sim$ 16 kHz) ASR models do not perform well for narrow band speech data \cite{389343}. The limited availability of labelled narrow band speech data is one of the biggest challenges in developing supervised ASR models for telephonic speech with good performance. 

A simple solution to tackle limited availability of narrow band speech data is to downsample the largely available wide band data and treat it as narrow band to build better performing narrow band ASR systems. But useful high frequency information is thrown away during downsampling \cite{389343,6424210}. Another approach is to perform bandwidth extension of narrow band speech, where the higher frequency components are generated using the correlation between low and high frequency components of the wide band speech signal to use ASR models built on wide band speech \cite{7178801,wang2015speech,li2015dnn}. In \cite{6424210}, it is shown that there is an improvement in performance with bandwidth extension for a small test set. Channel adaptation to narrow band speech with transfer learning is explored in \cite{zhuang2017improving}, where a model trained with the huge amount of wide band speech is finetuned on a small narrow band speech data. In \cite{mantena2019bandwidth}, a mixed bandwidth speech recognition model is explored in a supervised setting in which wide band speech is pooled with the narrow band speech to build a single model. This model gives improvement for both wide band and narrow band speech over the channel-specific models.  

Currently ASR models built on pretrained self-supervised models provide significant improvement in performance. Pretraining a self-supervised model requires huge amount of unlabelled data which is easy to obtain compared to labelled data. Robust \textit{Wav2Vec 2.0} \cite{hsu2021robust}, is a mixed bandwidth self-supervised model pretrained with data pooled from both narrow band (\textit{SwitchBoard}) and wide band speech. It gives improvements in downstream ASR for speech from both channels.

In this paper, a novel method is explored to improve the performance for both narrow and wide band speech with the state-of-the-art joint encoder-decoder self-supervised model \cite{arunkumar2022joint,ao2022pre,wu2022wav2seq}. This self-supervised training technique is an extension of \textit{HuBERT} \cite{hsu2021hubert}. \textit{HuBERT} tries to predict the acoustic class of the frames during training. In joint encoder-decoder self-supervised architecture, a decoder to the \textit{HuBERT} architecture and the target acoustic sequence is fed to the decoder after collapsing the repetitions during training. \cite{8639506} suggests that if the channel information is given to model by some means, the model will learn to differentiate it better and will give improvement in performance for speech from both the channels. So to train a channel-aware joint encoder-decoder self supervised model we propose to use non-overlapping acoustic units for the narrow and wide band speech. We then assess the performance of the trained mixed bandwidth joint encoder-decoder self-supervised model for both wide band and narrow band speech data.


The rest of the paper is organized as follows. Section \ref{sec:dataset} discusses the datasets and architecture used for the experiments. Section \ref{sec:widenarrow} shows the spectrum of wide and narrow band speech and explains the reasoning behind the proposed approach. Section \ref{sec:methodology} discusses the baseline mixed bandwidth joint encoder-decoder self-supervised model and the proposed method with novel non-overlaping acoustic units for the joint encoder-decodder self supervised model. In Section \ref{sec:expresults} the experimental results are discussed and finally the conclusions derived from the work are presented in Section \ref{sec:conclusion}.
\section{Datasets and Architecture Details}\label{sec:dataset}
\subsection{Datasets}
In this paper, four datasets of English language with different sampling rates are considered. The datasets included are:
\begin{enumerate}
\setlength\itemsep{0.5pt}
    \item \textit{LibriSpeech} \textit{train-clean-360} and \textit{train-clean-100} which is wide band speech with 16kHz sampling rate
    \item \textit{Switchboard - 100hr} and \textit{Switchboard - 300hr} which is telephonic narrow band speech with 8kHz sampling rate
    \item \textit{LibriSpeech} \textit{train-clean-100} downsampled at 8kHz to make pseudo data for read narrow band speech 
    \item \textit{TEDLIUM release2 - 100hr} subset down-sampled at 8KHz to make pseudo data for extempore  narrow band speech
\end{enumerate}
\subsection{Architecture Details}
The state-of-the-art Joint Encoder-Decoder SSL model is used for all the pretraining experiments \cite{arunkumar2022joint,ao2022pre,wu2022wav2seq}. A Transformer-decoder is incorporated with a conventional \textit{HuBERT} model. Like the transformer encoder-decoder architecture, the added decoder will attend to SSL encoder outputs through source attention. Target sequence of decoder is prepared from the discovered cluster IDs for all the encoder input frames. Consecutive repetitions in cluster IDs are replaced with single cluster ID so that the decoder can learn the sequence information. Details about the architecture and loss function are covered in detail in \cite{arunkumar2022joint}.

\section{Wide and Narrow Band Channel Spectral
Analysis}\label{sec:widenarrow}
\begin{figure}[h]
\graphicspath{ {images/} }
\centering
\begin{subfigure}[h]{0.48\columnwidth}
    \centering
    \includegraphics[width=\textwidth]{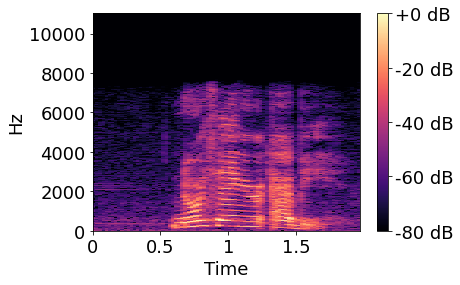}
    \caption{Sampling Rate = 16kHz}
    \label{16khz}
\end{subfigure}
\begin{subfigure}[h]{0.48\columnwidth}
    \centering
    \includegraphics[width=\textwidth]{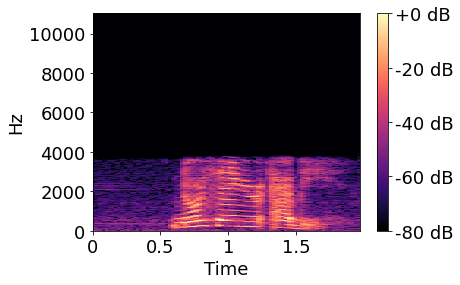}
    \caption{Sampling Rate = 8kHz}
    \label{8khz}
\end{subfigure}
\caption{Spectrogram of Wide and Narrow Band Speech Data}
\label{spectrogram}
\end{figure}

The spectrogram of wide and narrow speech signals are shown in Figure \ref{spectrogram}. A comparison of the two spectrograms of the same audio file sampled at 16kHz (Fig. \ref{16khz}) and 8kHz (Fig. \ref{8khz}) shows that the components of frequencies greater than 4kHz which are present in 16kHz sampled data are missing in 8kHz sampled data. A model trained on a pooled data with different sampling rate channels might not be able to learn the channel differences in the input. This can lead the model to erroneous predictions for the missing components of frequencies due to multi-channel dataset. 

\section{Methodology}\label{sec:methodology}
\subsection{Baseline}
The baseline for the proposed method is set up on the basis of three methods available in the literature.
\begin{itemize}
\setlength\itemsep{0.5pt}
    \item Pretraining the model using only wide band data
    \item Pretrainig  the model using only narrow band data
    \item Pretraining the model using both wide band and narrow band data pooled together
\end{itemize}
The details for the baseline experiment are as follows:
\subsubsection{Wide Band Pretraining : }
The model is pretrained only on \textit{LibriSpeech - 360hr} (16kHz dataset). Note that \textit{LibriSpeech - 360hr} is used despite having other larger corpora of \textit{LibriSpeech}. This justifies the comparison of models trained on wide and narrow band channels (\textit{Switchboard - 300hr}). 
\subsubsection{Narrow Band Pretraining : }
The model is pretrained on \textit{Switchboard - 300hr} which is narrow band in nature (8kHz). This is the only open source telephonic speech corpus currently available for the research purposes.
\subsubsection{Pooled Dataset Pretraining : }\label{pooled_dataset} 
\begin{figure}[h]
\graphicspath{ {images/} }
\centering
    \includegraphics[scale=0.5]{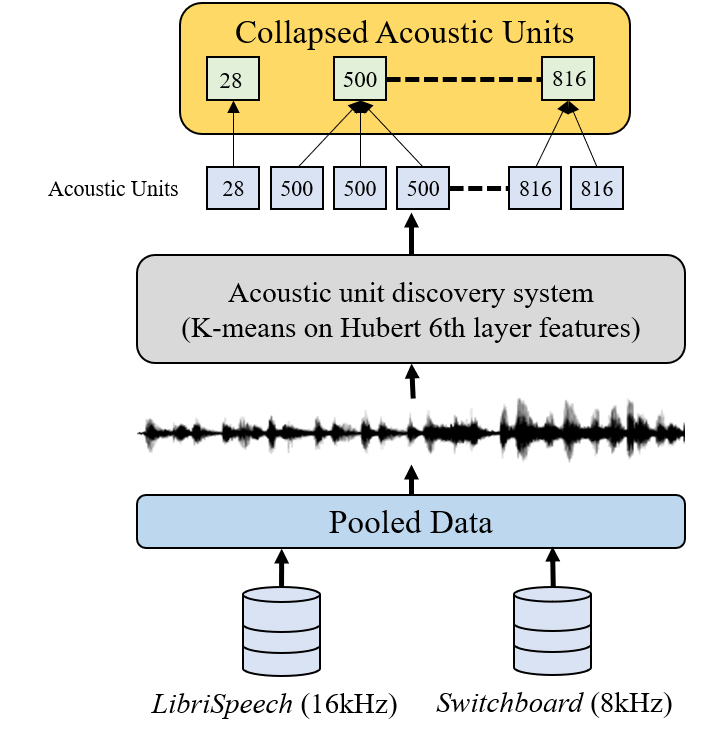}
    \caption{Pooled Dataset Clutering (Baseline Method)}
    \label{baseline}
\end{figure}

A reasonable approach to deal with individual disadvantages of exclusive pretraining on narrow and wide band dataset is to pool all the channel data of the same language. Since SSL pre-training is more dependent on the amount of data, pooling wide band data will also help in improving the performance for the narrow band data. This approach is used for the third baseline experiment. As shown in Figure \ref{baseline}, cluster ID corresponding to a frame is obtained from features extracted from \textit{HuBERT}'s sixth layer for the training data. The pooled dataset consists of the \textit{LibriSpeech train-clean-360} (wide band speech 16kHz) and \textit{Switchboard - 300hr} (narrow band speech 8kHz) datasets pre-training.

\begin{figure}[h]
    \centering  
    \includegraphics[width=\columnwidth]{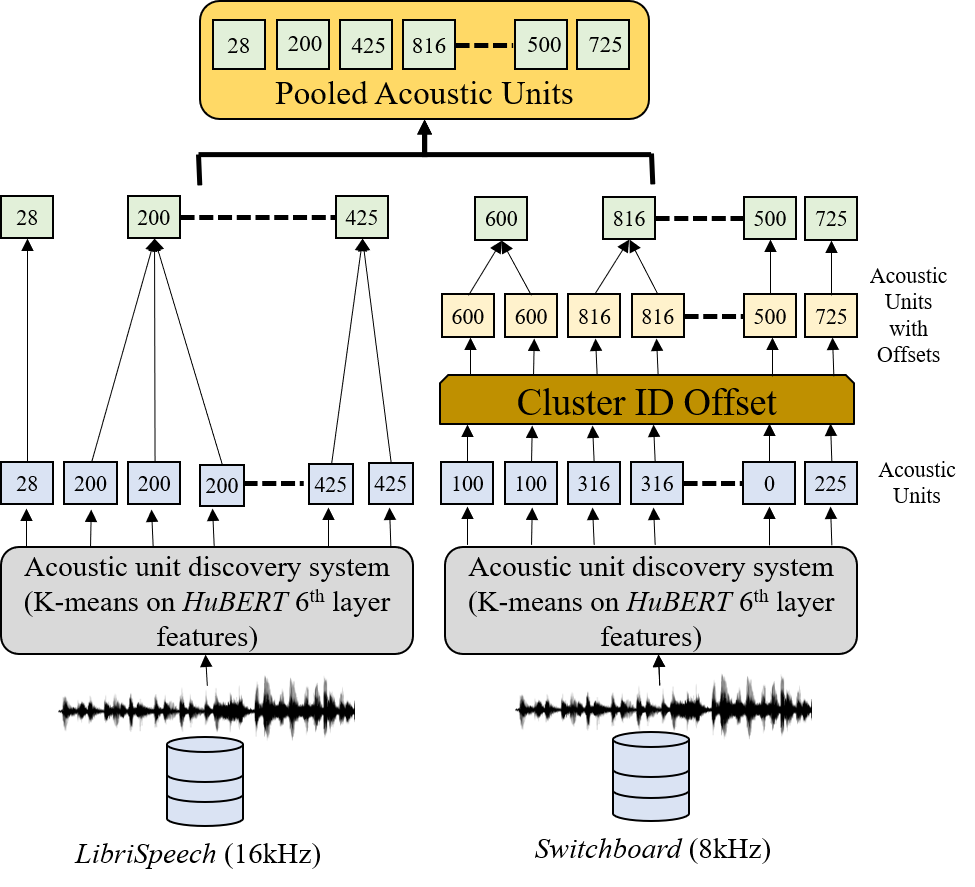}
    \caption{Proposed Method with Modified Clustering    (Number of Clusters 500 for Each Dataset and Offset=500)}
    \label{proposed}
    \vspace{-5pt}
\end{figure}
\subsection{Proposed Method}
As discussed in Section \ref{sec:widenarrow}, there is a difference in spectrogram of 8kHz and 16kHz speech signal which is not accounted for when a model is pretrained on a pooled dataset as discussed in Section \ref{pooled_dataset}. The idea behind the proposed method is that if the channel information is provided to the model, the model will start learning the channel difference better and will give the improvement over the baseline for narrow band speech data. Since acoustic unit sequence is taken as a target for decoder in joint encoder-decoder model, the channel information can be provided through the cluster ID. As shown in Figure \ref{proposed}, wide and narrow band data are clustered separately using k-means on \textit{HuBERT}'s $6^{th}$ layer features. A unique cluster ID is maintained before pooling the clusters from separate channels by offsetting the narrow band channel clusters by (at least) the number of clusters corresponding to wide band channel. 

\section{Experimental Results} \label{sec:expresults}
The hyper parameters for pretraining the joint encoder-decoder model are shown in Table \ref{table:sslmodel}.

\begin{table}[h]
\centering
\begin{tabular}{c c}
\hline
\textbf{Hyperparam-} & \textbf{Value} \\ 
\textbf{-eters} & \textbf{} \\ \hline \hline
Encoder layers           & 12             \\ \hline
Encoder units            & 3072           \\ \hline
Decoder layers           & 8              \\ \hline
Decoder units            & 2048           \\ \hline
\end{tabular}
\begin{tabular}{c c}
\hline
\textbf{Hyperparam-} & \textbf{Value} \\ 
\textbf{-eters} & \textbf{} \\ \hline \hline
Attention heads          & 8              \\ \hline
Batch-bins               & 15 Million     \\ \hline
Learning rate            & 0.0001         \\ \hline
Warmup steps             & 25000          \\ \hline
\end{tabular}
\caption{Hyperparamters for the Pretraining of SSL Model}
\label{table:sslmodel}
\end{table}
The number of clusters are 1000 for pooled dataset pretraining and for proposed method of modifying clustering algorithm, number of clusters are 500 for each narrow band and wide band dataset.

For ASR downstream task, both the encoder and decoder of the SSL model are fine-tuned with joint CTC/attention loss with $\beta = 0.3$ \cite{watanabe2018espnet}. The classification layers on top of the encoder and decoder are randomly initialized similar to \textit{HuBERT} finetuning and finetuned to predict character outputs. For ASR finetuning, the 3.2 million Batch-bins, 0.00002 Learning Rate, 8000 Warmup steps, and Adam optimizer are set as hyperparameters.
For all the experiments pretrained model is trained for 100 epochs. Pretrained models are finetuned for the narrow band and wide band speech data with the respective labelled data for 100 epochs. For testing the model, Word Error Rate matrix is used. 

For Tables \ref{narrow band result}, \ref{wide band results}, \ref{libri8k results}, and \ref{tedlium 8k}, please refer the following abbreviations:
\textbf{swbd300} : 300hr \textit{Switchboard} Unlabelled, \textbf{libri360} : 360hr \textit{LibriSpeech} Unlabelled, \textbf{swbd100} : 100hr \textit{Switchboard} labelled, \textbf{libri100} : 100hr \textit{LibriSpeech} labelled, \textbf{libri100-8k} : 100hr \textit{LibriSpeech} pseudo labelled, and \textbf{ted100-8k} : 100hr {\textit{TEDLIUM}} labelled. 
\subsection{Narrow Band Performance}
The pretrained models are finetuned on the \textit{Switchboard - 100hr} data to test the model on narrow band speech data. The results for the same are shown in Table \ref{narrow band result}.

\begin{table}[h]
\centering
\begin{tabular}{ccc}
\hline
{\begin{tabular}[c]{@{}c@{}}Pretraining\\Method\end{tabular}} & {Dataset}  &  {\begin{tabular}[c]{@{}c@{}}Swbd\\dev\end{tabular}} \\ \hline\hline
\begin{tabular}[c]{@{}c@{}}Wide \\ Band \end{tabular} &
\begin{tabular}[c]{@{}c@{}} swbd100 Labelled \\ \textbf{libri360 Unlabelled}\end{tabular}                                                & 17.8      \\ \hline
\begin{tabular}[c]{@{}c@{}}Narrow \\ Band \end{tabular} & 
\begin{tabular}[c]{@{}c@{}}swbd100 Labelled\\ \textbf{swbd300 Unlabelled}\end{tabular}                                                & 16.8      \\ \hline
\begin{tabular}[c]{@{}c@{}}Pooled\\Dataset \\(Baseline)\end{tabular} &
\begin{tabular}[c]{@{}c@{}}swbd100 Labelled\\ 
\textbf{660hr [swbd300}\\\textbf{+libri360 Unlabelled}\end{tabular}                            & 13.2      \\ \hline
\begin{tabular}[c]{@{}c@{}}Proposed with\\ modified\\clustering\end{tabular} &
\begin{tabular}[c]{@{}c@{}} swbd100 Labelled\\ 
\textbf{660hr [swbd300}\\\textbf{+libri300] Unlabelled}\end{tabular} & \textbf{12.6}  \\
\hline
\end{tabular}
\caption{Results for Narrow Band Test Dataset (WER)}
\label{narrow band result}
\end{table}

From Table \ref{narrow band result}, it can be seen that there is relative improvement of $\sim5.6\%$ for Narrow Band Pretraining compared with the Wide Band Pretraining. By giving the channel information while pretraining the model by modified clustering method, a relative improvement of $\sim5\%$ over the baseline (pooling the data without channel information) is achieved.


\subsection{Wide Band Performance}
The pretrained models are finetuned on the \textit{LibriSpeech - 100hr} data to test the model on wide band speech data. As shown in Table \ref{wide band results}, the wide band test data results show similar trends in improvements like narrow band results. The proposed method gives about $\sim4.3\%$ relative improvement in WER with respect to Pooled Dataset baseline. Hence, the proposed method improves performance for wide band as well as narrow band dataset.


\begin{table}[h]
\centering
\begin{tabular}{cccc}
\hline
{\begin{tabular}[c]{@{}c@{}}Pretraining\\Method\end{tabular}} & Dataset & \begin{tabular}[c]{@{}c@{}}test\\ clean\end{tabular} & \begin{tabular}[c]{@{}c@{}}test\\ other\end{tabular} \\ \hline\hline

\begin{tabular}[c]{@{}c@{}} Wide\\Band \end{tabular} &
\begin{tabular}[c]{@{}c@{}} libri100 Labelled\\ \textbf{libri360 Unlabelled}\end{tabular}                                                & 11.5                                                 & 30.6                                                 \\ \hline
\begin{tabular}[c]{@{}c@{}}Narrow \\ Band \end{tabular} & 
\begin{tabular}[c]{@{}c@{}}libri100 Labelled\\ \textbf{swbd300 Unlabelled}\end{tabular}                                                & 12.9    & 33.6  \\ \hline
\begin{tabular}[c]{@{}c@{}} Pooled\\Dataset\\ (Baseline) \end{tabular} &
\begin{tabular}[c]{@{}c@{}} libri100 Labelled\\ \textbf{660Hr [swbd300}\\\textbf{+libri360] Unlabelled}\end{tabular}                             & 8.2                                                  & 22.3                                                 \\ \hline
\begin{tabular}[c]{@{}c@{}} Proposed with\\modified\\clustering \end{tabular} &
\begin{tabular}[c]{@{}c@{}} libri100 Labelled\\ \textbf{660Hr [swbd300}\\\textbf{+libri360] Unlabelled}\end{tabular} & \textbf{7.6}                                                  & \textbf{21.5}\\                                      \hline          
\end{tabular}

\caption{Results for Wide Band Test Dataset (WER)}
\label{wide band results}
\end{table}
\vspace{-10pt}
\subsection{Performance on Pseudo Narrow band speech data}
To show the effect of pseudo  narrow band speech, all Pretrained models are finetuned and tested on \textit{LibriSpeech 100hr} and \textit{TEDLIUM release2 100hr}. 
Following observations can be made from the results of Table \ref{libri8k results} and \ref{tedlium 8k}
\begin{itemize}
    \item For \textit{LibriSpeech} 8KHz, the proposed method gives $\sim6.7\%$ and $\sim3.2\%$ relative improvement in WER for test-clean and test-other respectively, compared to pooled dataset baseline.
    \item For \textit{TEDLIUM} 8kHz, the proposed method gives $\sim4.0\%$ relative improvement for \textit{TEDLIUM} test dataset compared to pooled dataset baseline.
    \item Since \textit{Librispeech} is read speech data, the pseudo data matches wide band pretrained data and the wide band pretrained model gives better performance than the narrow band pretrained model.
    \item \textit{TEDLIUM} is extempore speech data which matches with narrow band pretrained data (telephonic speech). Hence, narrow band pretrained model gives better performance than the wide band pretrained model.
\end{itemize}

\begin{table}[h]
\centering
\begin{tabular}{cccc}
\hline
{\begin{tabular}[c]{@{}c@{}}Pretraining\\Method\end{tabular}} & Dataset & \begin{tabular}[c]{@{}c@{}}test\\ clean\end{tabular} & \begin{tabular}[c]{@{}c@{}}test\\ other\end{tabular} \\ \hline\hline

\begin{tabular}[c]{@{}c@{}} Wide\\Band \end{tabular} &
\begin{tabular}[c]{@{}c@{}}libri100-8k Labelled\\ \textbf{libri360 Unlabelled}\end{tabular}                                                & 13.2                                                 & 33.7                                                 \\ \hline
\begin{tabular}[c]{@{}c@{}}Narrow \\ Band \end{tabular} & 
\begin{tabular}[c]{@{}c@{}}libri100-8k Labelled\\ \textbf{swbd300 Unlabelled}\end{tabular}                                                & 14.2    & 34.2  \\ \hline

\begin{tabular}[c]{@{}c@{}} Pooled\\Dataset\\ (Baseline) \end{tabular} &
\begin{tabular}[c]{@{}c@{}} libri100-8k Labelled\\ \textbf{660Hr [swbd300}\\\textbf{+libri360] Unlabelled}\end{tabular}                             & 8.9                                                  & 24.3                                                 \\ \hline
\begin{tabular}[c]{@{}c@{}} Proposed with\\modified\\clustering \end{tabular} &
\begin{tabular}[c]{@{}c@{}} libri100-8k Labelled\\ \textbf{660Hr [swbd300}\\\textbf{+libri360] Unlabelled}\end{tabular} & \textbf{8.3}                                                  & \textbf{23.5}\\                                      \hline          
\end{tabular}

\caption{Results for Read Speech pseudo Narrow Band Test Dataset (WER) [Test sets and fine-tune data both are downsampled to 8KHz for this experiment.]}
\label{libri8k results}
\end{table}

\begin{table}[h]
\centering
\begin{tabular}{ccc}
\hline
{\begin{tabular}[c]{@{}c@{}}Pretraining\\Method\end{tabular}} & {Dataset}  &  {\begin{tabular}[c]{@{}c@{}}Swbd\\dev\end{tabular}} \\ \hline\hline
\begin{tabular}[c]{@{}c@{}}Wide \\ Band\end{tabular} &
\begin{tabular}[c]{@{}c@{}}ted100-8k Labelled \\ \textbf{libri360 Unlabelled}\end{tabular}                                                & 18.9      \\ \hline
\begin{tabular}[c]{@{}c@{}}Narrow \\ Band\end{tabular} & 
\begin{tabular}[c]{@{}c@{}}ted100-8k Labelled\\ \textbf{swbd300 Unlabelled}\end{tabular}                                                & 17.3      \\ \hline
\begin{tabular}[c]{@{}c@{}}Pooled\\Dataset\\(Baseline)\end{tabular} &
\begin{tabular}[c]{@{}c@{}}ted100-8k Labelled\\ 
\textbf{660hr [swbd300}\\\textbf{+libri360] Unlabelled}\end{tabular}                            & 12.4      \\ \hline
\begin{tabular}[c]{@{}c@{}}Proposed with\\ modified\\clustering\end{tabular} &
\begin{tabular}[c]{@{}c@{}}ted100-8k Labelled\\ 
\textbf{660hr [swbd300}\\\textbf{+libri360] Unlabelled}\end{tabular} & \textbf{11.9}  \\
\hline
\end{tabular}
\caption{Results for Extempore Pseudo Narrow Band Test Dataset (WER) [Test sets and fine-tune data both are downsampled to 8KHz for this experiment.]}
\label{tedlium 8k}
\end{table}
\vspace{-10pt}
\section{Conclusion} \label{sec:conclusion}

In this paper, we proposed a method for channel aware pertaining of the state-of-the-art joint encoder-decoder self-supervised architecture to handle the data pooled from different channels (narrow and wide band) in a better way. By giving different cluster IDs for the same speech units from different channels, the self-supervised model is able to differentiate the variation in same acoustic units coming from different sample rate. On an average a relative improvement of $\sim4\%$ is achieved over the selected self-supervised model trained with plain data pooling with different datasets. For future work this method can be tried on different accent or domain data.



\bibliography{ref}

\bibliographystyle{IEEEbib}

\end{document}